\documentclass{article}

\usepackage[dblblindworkshop, final]{neurips_2025}

\workshoptitle{MATH-AI}

\usepackage[utf8]{inputenc} 
\usepackage[T1]{fontenc}    
\usepackage[
    colorlinks,
    linkcolor={red!50!black},
    citecolor={blue!50!black},
    urlcolor={blue!80!black}]{hyperref}       
\usepackage{url}            
\usepackage{booktabs}       
\usepackage{amsfonts}       
\usepackage{nicefrac}       
\usepackage{microtype}      
\usepackage{xcolor}         
\usepackage{xspace}
\usepackage{booktabs}
\usepackage{todonotes}
\usepackage{cleveref}
\usepackage{subcaption}
\usepackage{enumitem}
\usepackage{minted}
\usepackage{array}
\usepackage{caption}
\usepackage{fontspec}
\usepackage{adjustbox}

\AddToHook{cmd/appendix/before}{%
    \crefalias{section}{appendix}%
    \crefalias{subsection}{appendix}
}

\usepackage{newfloat} 

\DeclareFloatingEnvironment[
    name=Example
]{example}

\usepackage{graphicx}
\usepackage{tikz}
\usetikzlibrary{fit, calc, patterns}

\usepackage[final]{listings}
\definecolor{keywordcolor}{rgb}{0.13, 0.54, 0.38}
\definecolor{tacticcolor}{rgb}{0.0, 0.1, 0.6}    
\definecolor{commentcolor}{rgb}{0.4, 0.4, 0.4}   
\definecolor{symbolcolor}{rgb}{0.0, 0.1, 0.6}    
\definecolor{sortcolor}{rgb}{0.1, 0.5, 0.1}      
\definecolor{attributecolor}{rgb}{0.7, 0.1, 0.1} 

\newcommand{\coqcommentstyle}{\color{blue}}

\lstdefinelanguage{coq}{
  mathescape=true,
  texcl=false,
  morekeywords=[1]{Section, Module, End, Require, Import, Export,
    Variable, Variables, Parameter, Parameters, Axiom, Hypothesis,
    Hypotheses, Notation, Local, Tactic, Reserved, Scope, Open, Close,
    Bind, Delimit, Definition, Equations, Let, Ltac, Fixpoint, CoFixpoint, Add,
    Morphism, Relation, Implicit, Arguments, Unset, Contextual,
    Strict, Prenex, Implicits, Inductive, CoInductive, Record,
    Structure, Canonical, Coercion, Context, Class, Global, Instance,
    Program, Infix, Theorem, Lemma, Corollary, Proposition, Fact, Fail,
    Remark, Example, Proof, Goal, Save, Qed, Defined, Hint, Resolve,
    Rewrite, View, Search, Show, Print, Printing, All, Eval, Check,
    Projections, inside, outside, Def},
  morekeywords=[2]{forall, exists, exists2, fun, fix, cofix, struct, by,
    match, with, end, as, in, return, let, if, is, then, else, for, of,
    nosimpl, when},
  morekeywords=[3]{Type, Prop, SProp, Set, true, false, option},
  morekeywords=[4]{pose, set, move, case, elim, apply, clear, hnf,
    intro, intros, generalize, rename, pattern, after, destruct,
    induction, using, refine, inversion, injection, rewrite, congr,
    unlock, compute, ring, field, fourier, replace, fold, unfold,
    change, cutrewrite, simpl, have, suff, wlog, suffices, without,
    loss, nat_norm, assert, cut, trivial, revert, bool_congr, nat_congr,
    symmetry, transitivity, auto, split, autorewrite},
  morekeywords=[5]{by, done, exact, reflexivity, tauto, romega, omega,
    assumption, solve, contradiction, discriminate},
  morekeywords=[6]{do, last, first, try, idtac, repeat},
  morecomment=[s]{(*}{*)},
  flexiblecolumns=false,
  showstringspaces=false,
  framesep=5pt,
  morestring=[b]",
  morestring=[d],
  tabsize=4,
  extendedchars=true,
  sensitive=true,
  breaklines=false,
  basicstyle=\ttfamily\small,
  captionpos=b,
  identifierstyle={\ttfamily\color{black}},
  keywordstyle=[1]{\ttfamily\bfseries\color{keywordcolor}},
  keywordstyle=[2]{\ttfamily\color{keywordcolor}},
  keywordstyle=[3]{\ttfamily\color{black}},
  keywordstyle=[4]{\ttfamily\color{VioletRed}},
  keywordstyle=[5]{\ttfamily\color{black}},
  keywordstyle=[6]{\ttfamily\color{black}},
  stringstyle=\ttfamily,
  commentstyle={\ttfamily\coqcommentstyle},
}[keywords,comments,strings]

\lstnewenvironment{coq}{\lstset{language=coq}}{}

\def\coqe{\lstinline[language=coq, basicstyle=\ttfamily]}

\newcommand{\rocq}{Rocq\xspace}
\newcommand{\petanque}{Petanque\xspace}

\newcommand{\pytanque}{\texttt{pytanque}\xspace}

\newcommand{\gpto}{4o mini\xspace}
\newcommand{\claude}{claude\xspace}
\newcommand{\oonemini}{o1 mini\xspace}
\newcommand{\oone}{o1\xspace}

\title{MiniF2F in \rocq: Automatic Translation Between Proof Assistants
    — A Case Study}

\author{%
  Jules Viennot \\
  IRIF, Université Paris Cité, Inria, CNRS \\
  \And
  Guillaume Baudart \\
  IRIF, Université Paris Cité, Inria, CNRS \\
  \And
  Emilio Jesús Gallego Arias \\
  IRIF, Université Paris Cité, CNRS \\
  \And
  Marc Lelarge \\
  DI ENS, PSL University, Inria \\
}

\begin{document}

\maketitle

\begin{abstract}
  While the MiniF2F dataset exists for Lean, Isabelle/HOL, and MetaMath, it has not been formalized in \rocq, limiting cross-system comparisons in automated theorem proving. We investigate whether state-of-the-art LLMs can automatically translate formal theorems between proof assistants. Using a three-stage methodology from basic prompting to multi-turn conversations with error feedback, we successfully translated 478 out of 488 theorems (98\%) from MiniF2F to \rocq. Expert validation of 150 translations confirmed high accuracy, with only three errors. This work provides a complete \rocq formalization of MiniF2F and demonstrates the viability of LLM-based cross-proof-assistant translation.
\end{abstract}


\section{Introduction}

Recent advances in Large Language Models (LLMs) have shown remarkable progress in automated theorem proving using interactive theorem provers (ITPs) such as Isabelle~\citep{DBLP:conf/nips/WuJLRSJS22,DBLP:journals/corr/abs-2303-04910}, Lean~\citep{DBLP:conf/iclr/PoluHZBBS23,DBLP:journals/corr/abs-2306-15626}, and \rocq~\citep{DBLP:journals/corr/abs-2305-04369,DBLP:journals/corr/abs-2412-14063}. However, the landscape of formal mathematics remains fragmented across different proof assistants, each with distinct syntactic conventions, type systems, and mathematical libraries. This fragmentation poses significant barriers to knowledge transfer and comparative evaluation.

The challenge of cross-system compatibility is particularly acute in the evaluation of machine learning approaches to theorem proving. Researchers developing techniques for different proof assistants often work with incompatible datasets, making it difficult to fairly compare methodologies or transfer insights across systems. While manual translation efforts exist, they are time-consuming, error-prone, and do not scale with the growing volume of formalized mathematics.

LLMs have demonstrated particular aptitude for translation tasks between programming languages, especially when extensive shared resources exist~\citep{DBLP:journals/corr/abs-2212-10079}. This capability suggests potential for automated translation between formal proof languages, which share many structural similarities despite their syntactic differences. Such automated translation could unlock significant value by enabling: (1) fair comparison of automated proving techniques across systems, (2) rapid porting of benchmark datasets, and (3) leveraging the unique strengths of different proof assistants for the same mathematical content.

In this work, we investigate whether state-of-the-art LLMs can effectively translate formal mathematical theorems between proof assistants. We focus specifically on translating the MiniF2F dataset~\citep{zheng2021minif2f,2210.12283} from its existing formalizations in Lean and Isabelle to \rocq. MiniF2F contains 488 high-school-level mathematical problems with existing formalizations in multiple systems, making it a popular benchmark for evaluating automated proof techniques~\citep{polu2020generative,thakur2024context,mikula2023magnushammer,wang2024proving}.

Our work is available at \url{https://github.com/LLM4Rocq/miniF2F-rocq}.

\section{Methodology}

We focus on translating MiniF2F to \rocq, as this system lacks a complete formalization despite previous community efforts.\footnote{\url{https://github.com/openai/miniF2F/issues/66}} The dataset contains 488 theorems spanning various mathematical domains including algebra, number theory, and geometry.

Our translation task generates \rocq theorem statements based on three input sources: (1) natural language descriptions of the mathematical problems, (2) existing Lean formalizations, and (3) existing Isabelle formalizations. We deliberately focus only on theorem statements, ignoring proofs to isolate the translation challenge from proof generation complexity.

All generated \rocq statements are automatically verified using \petanque and its dedicated interface for python \pytanque~\citep{nlir-mathai24}, a machine-to-machine interactive environment for \rocq. This ensures that our translations are both syntactically and type-theoretically correct within the \rocq system. Then, valid translations are reviewed by a human to ensure their semantic correctness with regards to the three input sources.


We designed a systematic approach with three stages of increasing complexity. To manage computational costs while maximizing translation success, we employ a cascading approach: each stage only processes theorems that remained untranslated in previous stages. This ensures that expensive model calls are focused on the most challenging cases while simpler theorems are handled efficiently in earlier stages.

\paragraph{Stage 1: one-shot prompting}

In this baseline stage, we provide models with a single prompt containing the natural language description and existing formalizations, requesting a direct \rocq translation. We evaluate four state-of-the-art models: GPT-4o mini (\gpto), Claude-3.5-Sonnet (\claude), o1-mini (\oonemini), and o1 (\oone). This stage assesses the models' inherent translation capabilities without interactive refinement.

\paragraph{Stage 2: multi-turn with error feedback}

Building on Stage 1 failures, we implement an interactive approach where models can attempt up to three translations per theorem. Each subsequent attempt incorporates the error messages from \petanque verification of previous attempts. This stage tests whether models can learn from their mistakes and iteratively improve translations. We focus on \claude and \oonemini for this stage based on their Stage 1 performance and cost considerations.

\paragraph{Stage 3: refined prompting with extended attempts}

For the most challenging remaining theorems, we implement targeted improvements using \claude. Based on error analysis from earlier stages, we refine our prompts to specifically address common failures.
We progressively increase the number of attempts from 6 to 24, allowing more extensive exploration of the solution space for difficult cases.

\section{Results}


\begin{figure}[h]
  \centering
  \scalebox{0.9}{\begin{tikzpicture}

    \draw[->, thick] (0,0) -- (9,0) node {};
    \draw[->, thick] (0,0) -- (0,5.7) node {};

    \foreach \y in {100, 200, 300, 400, 500} {
        \draw (-0.1, \y/100) -- (0.1, \y/100);
        \node[left] at (-0.2, \y/100) {\y};
    }

    \foreach \x/\label in {1/{4o mini}, 2/{claude}, 3/{\oonemini}, 4/{\oone}, 5/{claude(3)}, 6/{o1 mini(3)}, 7/{claude(6)}, 8/{claude(24)}} {
        \draw (\x, -0.1) -- (\x, 0.1);
        \node[rotate=45, left] at (\x, -0.2) {\footnotesize \label};
    }

    \draw (0.5,    0) -- (0.5, 1.93)
       -- (1.5, 1.93) -- (1.5, 2.73)
       -- (2.5, 2.73) -- (2.5, 2.87)
       -- (3.5, 2.87) -- (3.5, 3.32)
       -- (4.5, 3.32) -- (4.5, 3.80)
       -- (5.5, 3.80) -- (5.5, 3.94)
       -- (6.5, 3.94) -- (6.5, 4.65)
       -- (7.5, 4.65) -- (7.5, 4.78)
       -- (9, 4.78);

    \draw[dotted, red, thick] (0, 4.88) -- (9, 4.88);
    \draw[text=red, thick, draw=none, fill=white] (0.8, 4.9) circle (3mm) node {488};

    \foreach \x/\y in {1/193, 2/273, 3/287, 4/332, 5/380, 6/394, 7/465, 8/478} {
      \node[fit={(\x - 0.35, \y / 100 + 0.03) (\x + 0.35, \y / 100 + 0.3)}, inner sep=0pt, draw=none, fill=white, thick] (rect) {\footnotesize \y};
      \draw[dashed] (\x, 0) -- (\x, \y / 100);
    }

    \draw[blue, thick] (4.5, -0.1) -- (4.5, 5.5);
    \draw[blue, thick] (6.5, -0.1) -- (6.5, 5.5);

    \draw[text=blue, thick, draw=none, fill=white] (4, 5.3) circle (3mm) node {S1};
    \draw[text=blue, thick, draw=none, fill=white] (6, 5.3) circle (3mm) node {S2};
    \draw[text=blue, thick, draw=none, fill=white] (8.8, 5.3) circle (3mm) node {S3};

\end{tikzpicture}}
  \caption{Cumulative translation results for MiniF2F to \rocq across all experimental stages. Numbers in parentheses indicate the maximum number of attempts allowed per theorem in multi-turn stages.}
  \label{fig:results-minif2f}
\end{figure}

\Cref{fig:results-minif2f} presents our cumulative translation results across all stages. The progression demonstrates the value of our multi-stage approach.

One-shot prompting in Stage 1 achieved translation rates of up to 68\%, showing that models already possess strong base capabilities for translation between proof-assistants. Adding iterative attempts with error feedback in Stage 2 provided significant improvements: \claude successfully translated 31\% of the theorems remaining after Stage 1, demonstrating that models can learn from \rocq error messages. In Stage 3, refining the prompt and increasing the number of attempts yielded the most substantial gains, leaving only ten theorems untranslated. This highlights the importance of providing models with targeted information to overcome their limitations.

\paragraph{Quality assessment}

To validate translation quality, we conducted an expert audit on a random sample of 150 theorems (approximately 30\% of successful translations). Each expert reviewed a batch of~25 translations, comparing them against the natural language description as well as the Lean and Isabelle formalizations.

We classify the answers into three categories:

\begin{description}[font=\normalfont\itshape,leftmargin=1em,labelindent=1em,topsep=-0.3em, itemsep=0em]

  \item[Error:] the translation does not correspond to the original problem. For example, a translation with hypothesis \coqe|x| of type \coqe|Q| of a problem requiring the numerator and denominator of \coqe|x| to be relatively prime is false in \rocq, as this property is not guaranteed for elements of type \coqe|Q| (see also \Cref{sec:error}).

  \item[Perfectible:] the translation is correct but could be improved. For example, the \rocq statements \coqe|... (x > 0 /\ y > 0) -> ...| could be written \coqe|... x > 0 -> y > 0 -> ...| (see also \Cref{sec:perfectible}).

  \item[Valid:] the translation requires no modification.

\end{description}

\vspace*{-2mm}
\begin{table}[h]
  \caption{Expert audit results for 150 randomly sampled translations.}
  \label{tab:audit}
  \vspace*{3mm}
  \centering
  \begin{tabular}{@{}lc@{}}
    \toprule
    \textbf{Answers} & \textbf{Number of theorems} \\
    \midrule
    Error            & 3                           \\
    Perfectible      & 32                          \\
    Valid            & 115                         \\
    \bottomrule
  \end{tabular}
\end{table}

Results are presented in \Cref{tab:audit}.
The low error rate (2\%) and high rate of perfect translations (77\%) indicate that LLM-based translation can achieve human-level quality for the majority of cases.

\section{Discussion}

To better understand model capabilities for formal language translation, we now focus on four research questions:

\begin{description}
  \item[RQ1:] Do supposedly superior models actually perform better on translation tasks?
  \item[RQ2:] Does the amount of information available to the model affect its performance?
  \item[RQ3:] Is the generated code faithful to the existing formalizations?
  \item[RQ4:] Can a model assess the semantic correctness of translations?
\end{description}

\subsection{RQ1: models comparison}

To assess whether model rankings correlate with translation performance, we compared \gpto and \oonemini on a subset of 100 theorems. We evaluated both pass@1 (single attempt, equivalent to Stage 1) and pass@3 (three attempts) scenarios. As for Stage~1, a human ensured the semantic correctness of all proposed translations. Results are presented in \Cref{fig:results-rq1}.
\begin{figure}[ht]
  \centering
  \vspace{-0.5em}
  \begin{minipage}[t]{0.49\textwidth}
    \centering
    \scalebox{0.8}{\begin{tikzpicture}

    \def\sb{25};
    \def\ab{\sb / 100 * 360};

    \def\sfo{22};
    \def\afo{\sfo / 100 * 360};

    \def\som{10};
    \def\aom{\som / 100 * 360};

    \def\basea{3 * \afo / 4 + \ab / 2 + \som / 4 + 90};
    \def\baser{2};

    \fill[pattern=horizontal lines, pattern color=red]
        (0, 0) --
        ({\baser * cos(\basea)}, {\baser * sin(\basea)}) arc
        ({\basea}:{\basea - \afo - \ab}:{\baser}) --
        cycle;

    \fill[pattern=vertical lines, pattern color=blue]
        (0, 0) --
        ({\baser * cos(\basea - \afo)}, {\baser * sin(\basea - \afo)}) arc
        ({\basea - \afo}:{\basea - \afo - \ab - \aom}:{\baser}) --
        cycle;

    \draw (0, 0) circle (\baser);

    \node at ({\baser * 1.15 * cos(\basea - \afo / 2)},
              {\baser * 1.15 * sin(\basea - \afo / 2)}) {\sfo};
    \node at ({\baser * 1.15 * cos(\basea - \afo - \ab / 2)},
              {\baser * 1.15 * sin(\basea - \afo - \ab / 2)}) {\sb};
    \node at ({\baser * 1.15 * cos(\basea - \afo - \ab - \aom / 2)},
              {\baser * 1.15 * sin(\basea - \afo - \ab - \aom / 2)}) {\som};
    \node at (
        {\baser * 1.15 * cos(\basea - \afo / 2 - \ab / 2 - \aom / 2 - 180)},
        {\baser * 1.15 * sin(\basea - \afo / 2 - \ab / 2 - \aom / 2 - 180)}
        ) {43};
\end{tikzpicture}}
    \subcaption{pass@1}
  \end{minipage}
  \begin{minipage}[t]{0.49\textwidth}
    \centering
    \scalebox{0.8}{\begin{tikzpicture}

    \def\sb{54};
    \def\ab{\sb / 100 * 360};

    \def\sfo{7};
    \def\afo{\sfo / 100 * 360};

    \def\som{8};
    \def\aom{\som / 100 * 360};

    \def\basea{3 * \afo / 4 + \ab / 2 + \som / 4 + 90};
    \def\baser{2};

    \fill[pattern=horizontal lines, pattern color=red]
        (0, 0) --
        ({\baser * cos(\basea)}, {\baser * sin(\basea)}) arc
        ({\basea}:{\basea - \afo - \ab}:{\baser}) --
        cycle;

    \fill[pattern=vertical lines, pattern color=blue]
        (0, 0) --
        ({\baser * cos(\basea - \afo)}, {\baser * sin(\basea - \afo)}) arc
        ({\basea - \afo}:{\basea - \afo - \ab - \aom}:{\baser}) --
        cycle;

    \draw (0, 0) circle (\baser);

    \node at ({\baser * 1.15 * cos(\basea - \afo / 2)},
              {\baser * 1.15 * sin(\basea - \afo / 2)}) {\sfo};
    \node at ({\baser * 1.15 * cos(\basea - \afo - \ab / 2)},
              {\baser * 1.15 * sin(\basea - \afo - \ab / 2)}) {\sb};
    \node at ({\baser * 1.15 * cos(\basea - \afo - \ab - \aom / 2)},
              {\baser * 1.15 * sin(\basea - \afo - \ab - \aom / 2)}) {\som};
    \node at (
        {\baser * 1.15 * cos(\basea - \afo / 2 - \ab / 2 - \aom / 2 - 180)},
        {\baser * 1.15 * sin(\basea - \afo / 2 - \ab / 2 - \aom / 2 - 180)}
        ) {31};
\end{tikzpicture}}
    \subcaption{pass@3}
  \end{minipage}
  \caption{Comparison between \gpto and \oonemini performance. The circle area  amounts to the 100 theorems. Red horizontal lines denote theorems translated by \gpto. Blue vertical lines denote theorems translated by \oonemini. This defines four zones (untranslated, translated by \gpto only, translated by \oonemini only, translated by both models).}
  \label{fig:results-rq1}
\end{figure}

Despite \oonemini's chain-of-thought capabilities, \gpto achieved superior pass@1 performance. However, both models converged to similar performance levels at pass@3, suggesting that superior model architecture does not guarantee better translation performance.
Notably, both models tend to succeed and fail on the same theorems.

\subsection{RQ2: ablation study}

We conducted an ablation study using \oonemini on the same 100 theorems, systematically varying the input information:
informal description only, as it is the reference content on which the \rocq version must be based;
formal versions only, to test pure translation between proof assistants;
Lean version only as Lean is most similar to \rocq;
or everything at once (our initial set up).
The same methodology as in the models comparison is employed to compute pass@1 and pass@3 performance.

\vspace*{-2mm}
\begin{table}[h]
  \caption{Ablation study showing the effect of input information on translation.}
  \label{tab:ablation}
  \vspace*{3mm}
  \centering
  \begin{tabular}{@{}ccc@{}}
    \toprule
    \textbf{Information in the prompt} & \textbf{pass@1} & \textbf{pass@3} \\
    \midrule
    informal description + isabelle version + lean version & 35\% & 62\% \\
    \multicolumn{1}{@{}l}{informal description}            & 43\% & 65\% \\
    \multicolumn{1}{r}{isabelle version + lean version}    & 41\% & 62\% \\
    \multicolumn{1}{r}{lean version}                       & 40\% & 56\% \\
    \bottomrule
  \end{tabular}
\end{table}

Results are presented in \Cref{tab:ablation}.
Surprisingly, varying the input information does not substantially influence performance. Providing only the informal description achieved the best performance, suggesting that natural language descriptions constitute the most crucial information for models, while additional formal representations may introduce confusion rather than clarity.

\subsection{RQ3: faithfulness}

When evaluating semantic correctness of translations, we observed that a formalization can be valid for a theorem prover while failing to capture the complete intent of the natural language problem. For example, when expressing that $m$ is the maximum of a function $f$, formalizing only that $f$ is bounded above by $m$ is insufficient, the statement must also ensure that this bound is attained.

When both Lean and Isabelle formalizations were provided, mismatches originated from these reference versions (e.g., the maximum example above; see also \Cref{ex:adequacy}) in all but one case. This indicates that residual inaccuracies exist in the original MiniF2F formalizations, likely due to human error, and that our translated dataset achieves quality comparable to the original versions. To investigate these discrepancies, we analyzed results from the ablation study, focusing on the two prompting strategies containing informal descriptions: informal description only versus everything (Lean, Isabelle, and informal description).

\begin{figure}[ht]
  \centering
  \scalebox{0.9}{\begin{tikzpicture}[scale=0.8]

    \definecolor{faithcolor}{RGB}{255, 255, 255}
    \definecolor{validcolor}{RGB}{100, 100, 100}
    \definecolor{errcolor}{RGB}{0, 0, 0}

    \def\barwidth{0.6}
    \def\spacing{0.2}

    \draw[->] (0,0) -- (8.5,0) node[right] {};

    \foreach \x in {0, 10, 20, 30, 40, 50, 60, 70, 80} {
        \draw (\x/10, -0.1) -- (\x/10, 0);
        \node[below] at (\x/10, -0.2) {\footnotesize \x};
    }

    \draw (0, 0) -- (0, 9 * \spacing / 2 + 4 * \barwidth);
    \node[left] at (-0.2,     \spacing +     \barwidth / 2)
        {\footnotesize everything};
    \node[left] at (-0.2, 2 * \spacing + 3 * \barwidth / 2)
        {\footnotesize informal description};
    \node[left] at (-0.2, 3 * \spacing + 5 * \barwidth / 2)
        {\footnotesize everything};
    \node[left] at (-0.2, 4 * \spacing + 7 * \barwidth / 2)
        {\footnotesize informal description};

    \def\labeloffset{-4.2}
    \node[left] at (\labeloffset - 0.1, 7 * \spacing / 2 + 3 * \barwidth)
        {\footnotesize pass@1};
    \draw (\labeloffset, 3 * \spacing + 2 * \barwidth + 0.15)
       -- (\labeloffset, 4 * \spacing + 4 * \barwidth - 0.15);
    \node[left] at (\labeloffset - 0.1, 3 * \spacing / 2 +     \barwidth)
        {\footnotesize pass@3};
    \draw (\labeloffset,     \spacing +                 0.15)
       -- (\labeloffset, 2 * \spacing + 2 * \barwidth - 0.15);

    \fill[red, draw=red]
                  (4.3, 4 * \spacing + 3 * \barwidth)
        rectangle (5.1, 4 * \spacing + 4 * \barwidth);
    \fill[white, pattern color=red, pattern=crosshatch dots]
                  (3.8, 4 * \spacing + 3 * \barwidth)
        rectangle (4.3, 4 * \spacing + 4 * \barwidth);
    \draw[draw=black]
                  (0  , 4 * \spacing + 3 * \barwidth)
        rectangle (5.1, 4 * \spacing + 4 * \barwidth);

    \fill[red, draw=red]
                  (3.5, 3 * \spacing + 2 * \barwidth)
        rectangle (3.7, 3 * \spacing + 3 * \barwidth);
    \fill[white, pattern color=red, pattern=crosshatch dots]
                  (3.3, 3 * \spacing + 2 * \barwidth)
        rectangle (3.5, 3 * \spacing + 3 * \barwidth);
    \draw[draw=black]
                  (0  , 3 * \spacing + 2 * \barwidth)
        rectangle (3.7, 3 * \spacing + 3 * \barwidth);

    \fill[red, draw=red]
                  (6.5, 2 * \spacing +     \barwidth)
        rectangle (7.5, 2 * \spacing + 2 * \barwidth);
    \fill[white, pattern color=red, pattern=crosshatch dots]
                  (5.9, 2 * \spacing +     \barwidth)
        rectangle (6.5, 2 * \spacing + 2 * \barwidth);
    \draw[draw=black]
                  (0  , 2 * \spacing +     \barwidth)
        rectangle (7.5, 2 * \spacing + 2 * \barwidth);

    \fill[red, draw=red]
                  (6.2,     \spacing                )
        rectangle (6.4,     \spacing +     \barwidth);
    \fill[white, pattern color=red, pattern=crosshatch dots]
                  (5.8,     \spacing                )
        rectangle (6.2,     \spacing +     \barwidth);
    \draw[draw=black]
                  (0  ,     \spacing                )
        rectangle (6.4,     \spacing +     \barwidth);

\end{tikzpicture}}
  \caption{Effect of input information on translation faithfulness. Red bars represent errors, dotted bars indicate faithfulness errors, and white bars show valid translations.}
  \label{fig:faith}
\end{figure}

\Cref{fig:faith} reveals that prompting models with only informal descriptions produces more \rocq-accepted theorems but exhibits a higher overall error rate compared to including both Lean and Isabelle formalizations. Additionally, faithfulness errors constitute a larger proportion of total errors when formal versions are provided in the prompt.

\subsection{RQ4: LLM-as-a-judge}

To assess whether an LLM can perform semantic verification in place of human reviewers, we compared model judgments against human evaluations from the RQ1 and RQ2 experiments. For this task, we selected DeepSeek R1, a model distinct from those used for translation and known for strong reasoning capabilities. Detailed results are provided in \Cref{sec:report}.

DeepSeek R1 and human reviewers agreed on 95.2\% of translations. However, the model demonstrated limited accuracy in error detection: it failed to recognize 41.7\% of errors identified by human reviewers. Since the verification step aims to identify semantic errors, DeepSeek R1's poor error identification performance indicates that human review remains necessary for this task.

\section{Conclusion}

We successfully translated 478 of 488 theorems (98\%) from the MiniF2F dataset to \rocq using state-of-the-art LLMs, providing the first complete \rocq formalization of this important benchmark. Our three-stage methodology demonstrates that interactive approaches with error feedback substantially improve one-shot translation, with expert validation confirming high translation quality (only 2\% error rate). This work establishes LLM-based translation as a viable approach for translation between proof assistants.

\paragraph{Acknowledgements}
The research leading to these results has received funding from the Inria ``Défi" LLM4Code, and the European Research Council (ERC) under the European Union's Ninth Framework Programme Horizon Europe (ERC Synergy Project Malinca, Grant Agreement n.101167526).

\newpage

\bibliographystyle{unsrtnat}
\bibliography{minif2f}

\newpage

\appendix

\section{Appendix}
\label{sec:app}


\subsection{Example of an answer with an \textit{Error} and its correction}
\label{sec:error}

The natural language statement of the problem is: The number $a=\frac{p}{q}$, where $p$ and $q$ are relatively prime positive integers, has the property that the sum of all real numbers $x$ satisfying
$ \lfloor x \rfloor \cdot \{x\} = a \cdot x^2$
is $420$, where $\lfloor x \rfloor$ denotes the greatest integer less than or equal to $x$ and $\{x\}=x- \lfloor x \rfloor$ denotes the fractional part of $x$. What is $p+q$?\\
Show that it is 929.

\begin{example}[ht]
\begin{minipage}{0.49\textwidth}
\begin{lstlisting}[basicstyle=\footnotesize, language=coq]
Theorem amc12a_2020_p25 :
  forall (a : Q),

    forall (S : list R),
      (forall x : R, In x S <->
        (IZR (Int_part x) *
          (x - IZR (Int_part x))
          = Q2R a * Rpower x 2))
    -> NoDup S
    -> fold_left Rplus S 0 = 420
  -> (Z.pos (Qden a) + Qnum a = 929)%Z.
\end{lstlisting}
\subcaption{\rocq formalization before the audit.}
\end{minipage}
\begin{minipage}{0.49\textwidth}
\begin{lstlisting}[basicstyle=\footnotesize, language=coq]
Theorem amc12a_2020_p25 :
  forall (p q : nat),
      Nat.gcd p q = 1%nat ->
    forall (S : list R),
      (forall x : R, In x S <->
        (IZR (Int_part x) *
          (x - IZR (Int_part x))
          = INR p / INR q * Rpower x 2))
    -> NoDup S
    -> fold_left Rplus S 0 = 420
  -> (p + q = 929)%nat.
\end{lstlisting}
\subcaption{\rocq formalization after the audit.}
\end{minipage}
\caption{an answer with an \textit{Error} and its correction: \coqe|a| is replaced by the ratio of its numerator \coqe|p| and its denominator \coqe|q|, and a hypothesis ensuring they are relatively prime is added.}
\end{example}

\subsection{Example of a \textit{Perfectible} problem and its editing}
\label{sec:perfectible}
The natural language statement of the problem is: Let $a$ and $b$ be two positive real numbers, and $n$ be a positive integer.\\
Show that $(\frac{a+b}{2})^n \leq \frac{a^n+b^n}{2}$.

\begin{example}[ht]
\begin{minipage}{0.49\textwidth}
\begin{lstlisting}[basicstyle=\footnotesize, language=coq]
Theorem
  algebra_apbon2pownleqapownpbpowon2 :
  forall (a b : R) (n : nat),
    0 < a /\ 0 < b ->

    (0 < n)%nat ->
    ((a + b) / 2)^n
      <= (a ^ n + b ^ n) / 2.
\end{lstlisting}
\subcaption{\rocq formalization before the audit.}
\end{minipage}
\begin{minipage}{0.49\textwidth}
\begin{lstlisting}[basicstyle=\footnotesize, language=coq]
Theorem
  algebra_apbon2pownleqapownpbpowon2 :
  forall (a b : R) (n : nat),
    0 < a ->
    0 < b ->
    (0 < n)%nat ->
    ((a + b) / 2)^n
      <= (a ^ n + b ^ n) / 2.
\end{lstlisting}
\subcaption{\rocq formalization after the audit.}
\end{minipage}
\caption{a \textit{Perfectible} problem and its editing: the conjunction \coqe|a < 0 /\ b < 0| is curryfied into two separate hypotheses \coqe|a < 0| and \coqe|b < 0|.}
\end{example}

\subsection{Example of an answer that is not \emph{faithful} and its adjusment}\label{ex:adequacy}
The natural language statement of the problem is: \\
What is the maximum value of $\frac{(2^t-3t)t}{4^t}$ for real values of $t$?
Show that it is $\frac{1}{12}$.

In the Lean version\footnote{Lean formalization from \url{https://github.com/facebookresearch/miniF2F}}, only a proof for the upper bound is required:

\begin{lstlisting}[language=lean]
theorem amc12b_2020_p22
  (t : ℝ) :
  ((2^t - 3 * t) * t) / (4^t) ≤ 1 / 12 := ...
\end{lstlisting}

Consequently, the Rocq formalization returned by a model only requires to prove that $\frac{1}{12}$ is an upper bound. To align better with the informal statement, a statement ensuring the upper bound is reached is added:

\begin{lstlisting}[basicstyle=\footnotesize, language=coq]
Theorem amc12b_2020_p22 :
  forall t : R, ((exp (t * ln 2) - 3 * t) * t) / (exp (t * ln 4)) <= 1 / 12
  /\
  exists t : R, ((exp (t * ln 2) - 3 * t) * t) / (exp (t * ln 4)) = 1 / 12.
\end{lstlisting}

We consider it as an \emph{Faithfulness} issue: the formal statement in Lean or answered (before our review) is valid but not as strong as the informal statement.

\bigskip
Here is another example where the natural language statement is: Solve the system of equations
\begin{eqnarray*}
  |a_1 - a_2| x_2 +|a_1 - a_3| x_3 +|a_1 - a_4| x_4 &=& 1\\
  |a_2 - a_1| x_1 +|a_2 - a_3| x_3 +|a_2 - a_4| x_4 &=& 1\\
  |a_3 - a_1| x_1 +|a_3 - a_2| x_2 +|a_3-a_4|x_4 &=& 1\\
  |a_4 - a_1| x_1 +|a_4 - a_2| x_2 +|a_4 - a_3| x_3 &=& 1
\end{eqnarray*}
where $a_1, a_2, a_3, a_4$ are four different real numbers.

In this case, the formalization process requires to have a look at the solution. However, the informal proof (in \url{https://github.com/facebookresearch/miniF2F}) assumes that $a_1>a_2>a_3>a_4$ and shows that in this case, $x_2 = x_3 = 0$, and $x_1 = x_4 = 1/(a_1 - a_4)$. This informal proof only solves a particular case. It turns out that the general solution can be written as follows: define $m =\arg\max_i a_i$ and $n=\arg\min_i a_i$, then $x_m=x_n = \frac{1}{a_m-a_n}$ and for all $i\neq n,m$, $x_i =0$.

The Lean formalization relies on the weak informal proof and it is acknowledged in the file that this formal statetment is weaker than the informal original problem:
\begin{lstlisting}[language=lean]
-- Solution encoded in the theorem statement.
-- Conclusion too weak. It doesn't show "if and only if"
theorem imo_1966_p5 (x a : ℕ → ℝ) (h₀ : a 1 ≠ a 2) (h₁ : a 1 ≠ a 3) (h₂ : a 1 ≠ a 4)
  (h₃ : a 2 ≠ a 3) (h₄ : a 2 ≠ a 4) (h₅ : a 3 ≠ a 4) (h₆ : a 1 > a 2) (h₇ : a 2 > a 3)
  (h₈ : a 3 > a 4)
  (h₉ : abs (a 1 - a 2) * x 2 + abs (a 1 - a 3) * x 3 + abs (a 1 - a 4) * x 4 = 1)
  (h₁₀ : abs (a 2 - a 1) * x 1 + abs (a 2 - a 3) * x 3 + abs (a 2 - a 4) * x 4 = 1)
  (h₁₁ : abs (a 3 - a 1) * x 1 + abs (a 3 - a 2) * x 2 + abs (a 3 - a 4) * x 4 = 1)
  (h₁₂ : abs (a 4 - a 1) * x 1 + abs (a 4 - a 2) * x 2 + abs (a 4 - a 3) * x 3 = 1) :
  x 2 = 0 ∧ x 3 = 0 ∧ x 1 = 1 / abs (a 1 - a 4) ∧ x 4 = 1 / abs (a 1 - a 4) := by
  sorry
\end{lstlisting}

Thanks to our audit, we were able to get a stronger formal version closer to the original informal statement.
\begin{lstlisting}[basicstyle=\footnotesize, language=coq]
Theorem imo_1966_p5':
  forall (m n : nat) (x a : nat -> R),
  (forall i j, a i = a j -> i = j) ->
  (Rabs (a 1%nat - a 2%nat) * x 2%nat + Rabs (a 1%nat - a 3%nat) * x 3%nat
    + Rabs (a 1%nat - a 4%nat) * x 4%nat = INR 1) ->
  (Rabs (a 2%nat - a 1%nat) * x 1%nat + Rabs (a 2%nat - a 3%nat) * x 3%nat
    + Rabs (a 2%nat - a 4%nat) * x 4%nat = INR 1) ->
  (Rabs (a 3%nat - a 1%nat) * x 1%nat + Rabs (a 3%nat - a 2%nat) * x 2%nat
    + Rabs (a 3%nat - a 4%nat) * x 4%nat = INR 1) ->
  (Rabs (a 4%nat - a 1%nat) * x 1%nat + Rabs (a 4%nat - a 2%nat) * x 2%nat
    + Rabs (a 4%nat - a 3%nat) * x 3%nat = INR 1) ->
  (1 <= m <= 4 )%nat -> (1 <= n <= 4)%nat ->
  (forall i : nat, a m >= a i) ->
  (forall i, a n <= a i) ->
  (x m = 1/ (a m - a n)) /\ (x n = x m)
    /\ (forall i : nat , (i<=4)%nat -> i <> m -> i <> n -> x i = R0).
\end{lstlisting}


\subsection{Complete report of research questions results}
\label{sec:report}

Tables on the following pages present all detailed results computed in RQ1 and RQ2, listed by theorem.
All experimental configurations are represented: \gpto versus \oonemini comparisons, and various prompt information conditions for \oonemini.
For each configuration, we computed pass@3 results, generating three translations per theorem.
Within each table cell, results for the three translation attempts are presented and separated by blank spaces.
A dash indicates that the model failed to produce a valid \rocq statement.
For translations that successfully type-checked, results from the semantic verification phase are shown.
A \texttt{V} indicates a valid statement, a \texttt{F} denotes a faithfulness error, and a \texttt{E} represents an error.
Human reviews are displayed in black, while DeepSeek R1 answers are shown in gray.
The reviews where the assessments of humans and DeepSeek R1 diverged are highlighted in bold (\texttt{F} classifications by humans are considered valid for DeepSeek R1, as they align with the Lean and Isabelle formalizations in most cases).

\begin{table}[ht]
  \centering
  \makebox[\textwidth][c]{
  \begin{tabular}{@{}lccccc@{}}
  \toprule
  & \textbf{4o mini} & \multicolumn{4}{c}{\textbf{o1 mini}} \\
  \cmidrule(l{2mm}r{2mm}){2-2}
  \cmidrule(l{2mm}){3-6}
  \textbf{Theorems} & everything & everything & formal versions & lean version & informal description \\
  \midrule
  \footnotesize{\texttt{aime\_1983\_p1}} & \footnotesize{\texttt{-- -- V\textcolor{lightgray}{V}}} & \footnotesize{\texttt{-- -- --}} & \footnotesize{\texttt{-- -- --}} & \footnotesize{\texttt{-- -- --}} & \footnotesize{\texttt{-- -- --}} \\
  \footnotesize{\texttt{aime\_1990\_p4}} & \footnotesize{\texttt{-- V\textcolor{lightgray}{V} V\textcolor{lightgray}{V}}} & \footnotesize{\texttt{-- V\textcolor{lightgray}{V} --}} & \footnotesize{\texttt{V\textcolor{lightgray}{V} V\textcolor{lightgray}{V} V\textcolor{lightgray}{V}}} & \footnotesize{\texttt{V\textcolor{lightgray}{V} \textbf{V\textcolor{lightgray}{E}} --}} & \footnotesize{\texttt{E\textcolor{lightgray}{E} V\textcolor{lightgray}{V} V\textcolor{lightgray}{V}}} \\
  \footnotesize{\texttt{aime\_1991\_p6}} & \footnotesize{\texttt{-- -- --}} & \footnotesize{\texttt{-- -- --}} & \footnotesize{\texttt{-- -- --}} & \footnotesize{\texttt{-- -- --}} & \footnotesize{\texttt{-- -- --}} \\
  \footnotesize{\texttt{aimeII\_2001\_p3}} & \footnotesize{\texttt{-- -- --}} & \footnotesize{\texttt{-- -- --}} & \footnotesize{\texttt{-- -- --}} & \footnotesize{\texttt{-- -- --}} & \footnotesize{\texttt{-- -- --}} \\
  \footnotesize{\texttt{algebra\_3rootspoly\_...}} & \footnotesize{\texttt{-- -- V\textcolor{lightgray}{V}}} & \footnotesize{\texttt{V\textcolor{lightgray}{V} -- V\textcolor{lightgray}{V}}} & \footnotesize{\texttt{V\textcolor{lightgray}{V} V\textcolor{lightgray}{V} V\textcolor{lightgray}{V}}} & \footnotesize{\texttt{V\textcolor{lightgray}{V} V\textcolor{lightgray}{V} V\textcolor{lightgray}{V}}} & \footnotesize{\texttt{V\textcolor{lightgray}{V} -- V\textcolor{lightgray}{V}}} \\
  \footnotesize{\texttt{algebra\_9onxpypzleq...}} & \footnotesize{\texttt{V\textcolor{lightgray}{V} V\textcolor{lightgray}{V} V\textcolor{lightgray}{V}}} & \footnotesize{\texttt{V\textcolor{lightgray}{V} V\textcolor{lightgray}{V} --}} & \footnotesize{\texttt{-- V\textcolor{lightgray}{V} V\textcolor{lightgray}{V}}} & \footnotesize{\texttt{-- -- --}} & \footnotesize{\texttt{-- V\textcolor{lightgray}{V} V\textcolor{lightgray}{V}}} \\
  \footnotesize{\texttt{algebra\_apb4leq8ta4pb4}} & \footnotesize{\texttt{V\textcolor{lightgray}{V} V\textcolor{lightgray}{V} V\textcolor{lightgray}{V}}} & \footnotesize{\texttt{V\textcolor{lightgray}{V} V\textcolor{lightgray}{V} V\textcolor{lightgray}{V}}} & \footnotesize{\texttt{V\textcolor{lightgray}{V} -- V\textcolor{lightgray}{V}}} & \footnotesize{\texttt{-- V\textcolor{lightgray}{V} V\textcolor{lightgray}{V}}} & \footnotesize{\texttt{V\textcolor{lightgray}{V} V\textcolor{lightgray}{V} V\textcolor{lightgray}{V}}} \\
  \footnotesize{\texttt{algebra\_others\_exir...}} & \footnotesize{\texttt{-- -- --}} & \footnotesize{\texttt{-- -- --}} & \footnotesize{\texttt{-- -- --}} & \footnotesize{\texttt{E\textcolor{lightgray}{E} -- --}} & \footnotesize{\texttt{-- -- --}} \\
  \footnotesize{\texttt{algebra\_sqineq\_at2m...}} & \footnotesize{\texttt{V\textcolor{lightgray}{V} V\textcolor{lightgray}{V} V\textcolor{lightgray}{V}}} & \footnotesize{\texttt{V\textcolor{lightgray}{V} V\textcolor{lightgray}{V} V\textcolor{lightgray}{V}}} & \footnotesize{\texttt{V\textcolor{lightgray}{V} -- V\textcolor{lightgray}{V}}} & \footnotesize{\texttt{V\textcolor{lightgray}{V} V\textcolor{lightgray}{V} V\textcolor{lightgray}{V}}} & \footnotesize{\texttt{V\textcolor{lightgray}{V} V\textcolor{lightgray}{V} V\textcolor{lightgray}{V}}} \\
  \footnotesize{\texttt{amc12\_2000\_p6}} & \footnotesize{\texttt{-- -- --}} & \footnotesize{\texttt{-- -- --}} & \footnotesize{\texttt{-- -- --}} & \footnotesize{\texttt{-- -- --}} & \footnotesize{\texttt{-- -- --}} \\
  \footnotesize{\texttt{amc12\_2001\_p9}} & \footnotesize{\texttt{V\textcolor{lightgray}{V} -- V\textcolor{lightgray}{V}}} & \footnotesize{\texttt{V\textcolor{lightgray}{V} V\textcolor{lightgray}{V} V\textcolor{lightgray}{V}}} & \footnotesize{\texttt{V\textcolor{lightgray}{V} V\textcolor{lightgray}{V} V\textcolor{lightgray}{V}}} & \footnotesize{\texttt{-- -- V\textcolor{lightgray}{V}}} & \footnotesize{\texttt{V\textcolor{lightgray}{V} V\textcolor{lightgray}{V} V\textcolor{lightgray}{V}}} \\
  \footnotesize{\texttt{amc12a\_2002\_p21}} & \footnotesize{\texttt{F\textcolor{lightgray}{V} -- --}} & \footnotesize{\texttt{-- -- --}} & \footnotesize{\texttt{\textbf{F\textcolor{lightgray}{E}} \textbf{F\textcolor{lightgray}{E}} \textbf{F\textcolor{lightgray}{E}}}} & \footnotesize{\texttt{F\textcolor{lightgray}{V} -- --}} & \footnotesize{\texttt{-- -- --}} \\
  \footnotesize{\texttt{amc12a\_2003\_p5}} & \footnotesize{\texttt{-- -- E\textcolor{lightgray}{E}}} & \footnotesize{\texttt{V\textcolor{lightgray}{V} V\textcolor{lightgray}{V} V\textcolor{lightgray}{V}}} & \footnotesize{\texttt{-- -- --}} & \footnotesize{\texttt{\textbf{E\textcolor{lightgray}{V}} -- --}} & \footnotesize{\texttt{-- -- \textbf{E\textcolor{lightgray}{V}}}} \\
  \footnotesize{\texttt{amc12a\_2008\_p15}} & \footnotesize{\texttt{V\textcolor{lightgray}{V} V\textcolor{lightgray}{V} V\textcolor{lightgray}{V}}} & \footnotesize{\texttt{-- V\textcolor{lightgray}{V} V\textcolor{lightgray}{V}}} & \footnotesize{\texttt{V\textcolor{lightgray}{V} -- V\textcolor{lightgray}{V}}} & \footnotesize{\texttt{V\textcolor{lightgray}{V} V\textcolor{lightgray}{V} V\textcolor{lightgray}{V}}} & \footnotesize{\texttt{V\textcolor{lightgray}{V} -- V\textcolor{lightgray}{V}}} \\
  \footnotesize{\texttt{amc12a\_2009\_p2}} & \footnotesize{\texttt{\textbf{E\textcolor{lightgray}{V}} -- V\textcolor{lightgray}{V}}} & \footnotesize{\texttt{-- V\textcolor{lightgray}{V} V\textcolor{lightgray}{V}}} & \footnotesize{\texttt{V\textcolor{lightgray}{V} -- V\textcolor{lightgray}{V}}} & \footnotesize{\texttt{V\textcolor{lightgray}{V} V\textcolor{lightgray}{V} \textbf{E\textcolor{lightgray}{V}}}} & \footnotesize{\texttt{V\textcolor{lightgray}{V} V\textcolor{lightgray}{V} V\textcolor{lightgray}{V}}} \\
  \footnotesize{\texttt{amc12a\_2009\_p9}} & \footnotesize{\texttt{V\textcolor{lightgray}{V} V\textcolor{lightgray}{V} V\textcolor{lightgray}{V}}} & \footnotesize{\texttt{-- -- --}} & \footnotesize{\texttt{V\textcolor{lightgray}{V} -- --}} & \footnotesize{\texttt{V\textcolor{lightgray}{V} -- --}} & \footnotesize{\texttt{V\textcolor{lightgray}{V} V\textcolor{lightgray}{V} V\textcolor{lightgray}{V}}} \\
  \footnotesize{\texttt{amc12a\_2010\_p11}} & \footnotesize{\texttt{-- -- --}} & \footnotesize{\texttt{-- -- --}} & \footnotesize{\texttt{-- -- V\textcolor{lightgray}{V}}} & \footnotesize{\texttt{-- -- --}} & \footnotesize{\texttt{-- -- --}} \\
  \footnotesize{\texttt{amc12a\_2013\_p7}} & \footnotesize{\texttt{-- -- --}} & \footnotesize{\texttt{-- -- --}} & \footnotesize{\texttt{-- -- --}} & \footnotesize{\texttt{-- -- --}} & \footnotesize{\texttt{-- -- V\textcolor{lightgray}{V}}} \\
  \footnotesize{\texttt{amc12a\_2013\_p8}} & \footnotesize{\texttt{V\textcolor{lightgray}{V} -- --}} & \footnotesize{\texttt{V\textcolor{lightgray}{V} V\textcolor{lightgray}{V} V\textcolor{lightgray}{V}}} & \footnotesize{\texttt{-- -- --}} & \footnotesize{\texttt{-- V\textcolor{lightgray}{V} --}} & \footnotesize{\texttt{-- V\textcolor{lightgray}{V} V\textcolor{lightgray}{V}}} \\
  \footnotesize{\texttt{amc12a\_2017\_p2}} & \footnotesize{\texttt{V\textcolor{lightgray}{V} V\textcolor{lightgray}{V} V\textcolor{lightgray}{V}}} & \footnotesize{\texttt{V\textcolor{lightgray}{V} V\textcolor{lightgray}{V} --}} & \footnotesize{\texttt{V\textcolor{lightgray}{V} V\textcolor{lightgray}{V} V\textcolor{lightgray}{V}}} & \footnotesize{\texttt{-- -- --}} & \footnotesize{\texttt{-- V\textcolor{lightgray}{V} V\textcolor{lightgray}{V}}} \\
  \footnotesize{\texttt{amc12a\_2020\_p21}} & \footnotesize{\texttt{-- -- --}} & \footnotesize{\texttt{-- -- --}} & \footnotesize{\texttt{-- -- --}} & \footnotesize{\texttt{-- -- --}} & \footnotesize{\texttt{-- -- --}} \\
  \footnotesize{\texttt{amc12a\_2020\_p25}} & \footnotesize{\texttt{-- -- --}} & \footnotesize{\texttt{-- -- --}} & \footnotesize{\texttt{-- -- --}} & \footnotesize{\texttt{-- -- --}} & \footnotesize{\texttt{-- -- --}} \\
  \footnotesize{\texttt{amc12a\_2021\_p8}} & \footnotesize{\texttt{-- -- --}} & \footnotesize{\texttt{-- -- V\textcolor{lightgray}{V}}} & \footnotesize{\texttt{-- -- --}} & \footnotesize{\texttt{-- -- --}} & \footnotesize{\texttt{-- E\textcolor{lightgray}{E} --}} \\
  \footnotesize{\texttt{amc12b\_2003\_p9}} & \footnotesize{\texttt{-- V\textcolor{lightgray}{V} V\textcolor{lightgray}{V}}} & \footnotesize{\texttt{-- V\textcolor{lightgray}{V} V\textcolor{lightgray}{V}}} & \footnotesize{\texttt{-- -- V\textcolor{lightgray}{V}}} & \footnotesize{\texttt{-- -- --}} & \footnotesize{\texttt{V\textcolor{lightgray}{V} -- --}} \\
  \footnotesize{\texttt{amc12b\_2004\_p3}} & \footnotesize{\texttt{V\textcolor{lightgray}{V} V\textcolor{lightgray}{V} V\textcolor{lightgray}{V}}} & \footnotesize{\texttt{V\textcolor{lightgray}{V} V\textcolor{lightgray}{V} V\textcolor{lightgray}{V}}} & \footnotesize{\texttt{-- V\textcolor{lightgray}{V} V\textcolor{lightgray}{V}}} & \footnotesize{\texttt{V\textcolor{lightgray}{V} -- V\textcolor{lightgray}{V}}} & \footnotesize{\texttt{-- V\textcolor{lightgray}{V} V\textcolor{lightgray}{V}}} \\
  \footnotesize{\texttt{amc12b\_2020\_p13}} & \footnotesize{\texttt{-- -- --}} & \footnotesize{\texttt{-- -- --}} & \footnotesize{\texttt{-- -- --}} & \footnotesize{\texttt{-- V\textcolor{lightgray}{V} --}} & \footnotesize{\texttt{-- -- --}} \\
  \footnotesize{\texttt{amc12b\_2020\_p22}} & \footnotesize{\texttt{-- -- F\textcolor{lightgray}{V}}} & \footnotesize{\texttt{-- F\textcolor{lightgray}{V} --}} & \footnotesize{\texttt{-- -- --}} & \footnotesize{\texttt{-- -- --}} & \footnotesize{\texttt{V\textcolor{lightgray}{V} -- --}} \\
  \footnotesize{\texttt{amc12b\_2021\_p18}} & \footnotesize{\texttt{-- -- --}} & \footnotesize{\texttt{-- -- --}} & \footnotesize{\texttt{-- V\textcolor{lightgray}{V} V\textcolor{lightgray}{V}}} & \footnotesize{\texttt{-- -- --}} & \footnotesize{\texttt{-- -- --}} \\
  \footnotesize{\texttt{imo\_1977\_p5}} & \footnotesize{\texttt{-- -- --}} & \footnotesize{\texttt{-- -- --}} & \footnotesize{\texttt{V\textcolor{lightgray}{V} -- --}} & \footnotesize{\texttt{-- -- --}} & \footnotesize{\texttt{\textbf{E\textcolor{lightgray}{V}} -- E\textcolor{lightgray}{E}}} \\
  \footnotesize{\texttt{imo\_1977\_p6}} & \footnotesize{\texttt{V\textcolor{lightgray}{V} -- V\textcolor{lightgray}{V}}} & \footnotesize{\texttt{V\textcolor{lightgray}{V} V\textcolor{lightgray}{V} V\textcolor{lightgray}{V}}} & \footnotesize{\texttt{-- -- V\textcolor{lightgray}{V}}} & \footnotesize{\texttt{V\textcolor{lightgray}{V} -- --}} & \footnotesize{\texttt{E\textcolor{lightgray}{E} E\textcolor{lightgray}{E} E\textcolor{lightgray}{E}}} \\
  \footnotesize{\texttt{imo\_1981\_p6}} & \footnotesize{\texttt{F\textcolor{lightgray}{V} F\textcolor{lightgray}{V} --}} & \footnotesize{\texttt{-- F\textcolor{lightgray}{V} --}} & \footnotesize{\texttt{V\textcolor{lightgray}{V} F\textcolor{lightgray}{V} F\textcolor{lightgray}{V}}} & \footnotesize{\texttt{F\textcolor{lightgray}{V} F\textcolor{lightgray}{V} --}} & \footnotesize{\texttt{-- -- --}} \\
  \footnotesize{\texttt{imo\_1997\_p5}} & \footnotesize{\texttt{V\textcolor{lightgray}{V} -- V\textcolor{lightgray}{V}}} & \footnotesize{\texttt{V\textcolor{lightgray}{V} -- --}} & \footnotesize{\texttt{V\textcolor{lightgray}{V} -- --}} & \footnotesize{\texttt{V\textcolor{lightgray}{V} -- V\textcolor{lightgray}{V}}} & \footnotesize{\texttt{V\textcolor{lightgray}{V} -- V\textcolor{lightgray}{V}}} \\
  \footnotesize{\texttt{imo\_2001\_p6}} & \footnotesize{\texttt{-- -- --}} & \footnotesize{\texttt{-- -- --}} & \footnotesize{\texttt{-- -- --}} & \footnotesize{\texttt{-- -- --}} & \footnotesize{\texttt{-- -- --}} \\
  \footnotesize{\texttt{imosl\_2007\_algebra\_p6}} & \footnotesize{\texttt{-- -- --}} & \footnotesize{\texttt{-- -- --}} & \footnotesize{\texttt{-- -- --}} & \footnotesize{\texttt{-- -- --}} & \footnotesize{\texttt{-- -- --}} \\
  \footnotesize{\texttt{induction\_nfactltne...}} & \footnotesize{\texttt{-- -- --}} & \footnotesize{\texttt{-- V\textcolor{lightgray}{V} V\textcolor{lightgray}{V}}} & \footnotesize{\texttt{-- V\textcolor{lightgray}{V} --}} & \footnotesize{\texttt{-- -- --}} & \footnotesize{\texttt{-- -- V\textcolor{lightgray}{V}}} \\
  \footnotesize{\texttt{induction\_seq\_mul2pnp1}} & \footnotesize{\texttt{V\textcolor{lightgray}{V} V\textcolor{lightgray}{V} V\textcolor{lightgray}{V}}} & \footnotesize{\texttt{-- V\textcolor{lightgray}{V} V\textcolor{lightgray}{V}}} & \footnotesize{\texttt{-- -- --}} & \footnotesize{\texttt{-- -- --}} & \footnotesize{\texttt{V\textcolor{lightgray}{V} -- --}} \\
  \footnotesize{\texttt{induction\_sum\_1oktkp1}} & \footnotesize{\texttt{-- -- --}} & \footnotesize{\texttt{-- -- --}} & \footnotesize{\texttt{-- -- --}} & \footnotesize{\texttt{-- -- V\textcolor{lightgray}{V}}} & \footnotesize{\texttt{-- -- --}} \\
  \footnotesize{\texttt{mathd\_algebra\_13}} & \footnotesize{\texttt{V\textcolor{lightgray}{V} V\textcolor{lightgray}{V} V\textcolor{lightgray}{V}}} & \footnotesize{\texttt{-- V\textcolor{lightgray}{V} V\textcolor{lightgray}{V}}} & \footnotesize{\texttt{V\textcolor{lightgray}{V} -- --}} & \footnotesize{\texttt{V\textcolor{lightgray}{V} -- --}} & \footnotesize{\texttt{V\textcolor{lightgray}{V} V\textcolor{lightgray}{V} E\textcolor{lightgray}{E}}} \\
  \footnotesize{\texttt{mathd\_algebra\_15}} & \footnotesize{\texttt{V\textcolor{lightgray}{V} -- V\textcolor{lightgray}{V}}} & \footnotesize{\texttt{-- V\textcolor{lightgray}{V} V\textcolor{lightgray}{V}}} & \footnotesize{\texttt{-- -- --}} & \footnotesize{\texttt{V\textcolor{lightgray}{V} -- --}} & \footnotesize{\texttt{V\textcolor{lightgray}{V} -- --}} \\
  \footnotesize{\texttt{mathd\_algebra\_24}} & \footnotesize{\texttt{V\textcolor{lightgray}{V} V\textcolor{lightgray}{V} V\textcolor{lightgray}{V}}} & \footnotesize{\texttt{-- V\textcolor{lightgray}{V} V\textcolor{lightgray}{V}}} & \footnotesize{\texttt{V\textcolor{lightgray}{V} V\textcolor{lightgray}{V} V\textcolor{lightgray}{V}}} & \footnotesize{\texttt{V\textcolor{lightgray}{V} V\textcolor{lightgray}{V} V\textcolor{lightgray}{V}}} & \footnotesize{\texttt{V\textcolor{lightgray}{V} V\textcolor{lightgray}{V} V\textcolor{lightgray}{V}}} \\
  \footnotesize{\texttt{mathd\_algebra\_48}} & \footnotesize{\texttt{-- V\textcolor{lightgray}{V} V\textcolor{lightgray}{V}}} & \footnotesize{\texttt{-- V\textcolor{lightgray}{V} V\textcolor{lightgray}{V}}} & \footnotesize{\texttt{V\textcolor{lightgray}{V} V\textcolor{lightgray}{V} \textbf{V\textcolor{lightgray}{E}}}} & \footnotesize{\texttt{V\textcolor{lightgray}{V} \textbf{V\textcolor{lightgray}{E}} V\textcolor{lightgray}{V}}} & \footnotesize{\texttt{V\textcolor{lightgray}{V} V\textcolor{lightgray}{V} --}} \\
  \footnotesize{\texttt{mathd\_algebra\_51}} & \footnotesize{\texttt{V\textcolor{lightgray}{V} V\textcolor{lightgray}{V} V\textcolor{lightgray}{V}}} & \footnotesize{\texttt{-- -- --}} & \footnotesize{\texttt{-- V\textcolor{lightgray}{V} V\textcolor{lightgray}{V}}} & \footnotesize{\texttt{V\textcolor{lightgray}{V} -- V\textcolor{lightgray}{V}}} & \footnotesize{\texttt{V\textcolor{lightgray}{V} V\textcolor{lightgray}{V} V\textcolor{lightgray}{V}}} \\
  \footnotesize{\texttt{mathd\_algebra\_67}} & \footnotesize{\texttt{V\textcolor{lightgray}{V} V\textcolor{lightgray}{V} V\textcolor{lightgray}{V}}} & \footnotesize{\texttt{V\textcolor{lightgray}{V} V\textcolor{lightgray}{V} --}} & \footnotesize{\texttt{V\textcolor{lightgray}{V} -- V\textcolor{lightgray}{V}}} & \footnotesize{\texttt{V\textcolor{lightgray}{V} -- --}} & \footnotesize{\texttt{V\textcolor{lightgray}{V} -- --}} \\
  \footnotesize{\texttt{mathd\_algebra\_77}} & \footnotesize{\texttt{V\textcolor{lightgray}{V} V\textcolor{lightgray}{V} --}} & \footnotesize{\texttt{-- V\textcolor{lightgray}{V} V\textcolor{lightgray}{V}}} & \footnotesize{\texttt{V\textcolor{lightgray}{V} V\textcolor{lightgray}{V} --}} & \footnotesize{\texttt{-- -- V\textcolor{lightgray}{V}}} & \footnotesize{\texttt{V\textcolor{lightgray}{V} V\textcolor{lightgray}{V} V\textcolor{lightgray}{V}}} \\
  \footnotesize{\texttt{mathd\_algebra\_104}} & \footnotesize{\texttt{V\textcolor{lightgray}{V} V\textcolor{lightgray}{V} V\textcolor{lightgray}{V}}} & \footnotesize{\texttt{V\textcolor{lightgray}{V} -- V\textcolor{lightgray}{V}}} & \footnotesize{\texttt{V\textcolor{lightgray}{V} V\textcolor{lightgray}{V} V\textcolor{lightgray}{V}}} & \footnotesize{\texttt{V\textcolor{lightgray}{V} \textbf{V\textcolor{lightgray}{E}} V\textcolor{lightgray}{V}}} & \footnotesize{\texttt{V\textcolor{lightgray}{V} V\textcolor{lightgray}{V} V\textcolor{lightgray}{V}}} \\
  \footnotesize{\texttt{mathd\_algebra\_107}} & \footnotesize{\texttt{F\textcolor{lightgray}{V} F\textcolor{lightgray}{V} F\textcolor{lightgray}{V}}} & \footnotesize{\texttt{V\textcolor{lightgray}{V} F\textcolor{lightgray}{V} F\textcolor{lightgray}{V}}} & \footnotesize{\texttt{F\textcolor{lightgray}{V} F\textcolor{lightgray}{V} --}} & \footnotesize{\texttt{F\textcolor{lightgray}{V} F\textcolor{lightgray}{V} \textbf{F\textcolor{lightgray}{E}}}} & \footnotesize{\texttt{F\textcolor{lightgray}{V} F\textcolor{lightgray}{V} \textbf{E\textcolor{lightgray}{V}}}} \\
  \footnotesize{\texttt{mathd\_algebra\_119}} & \footnotesize{\texttt{V\textcolor{lightgray}{V} V\textcolor{lightgray}{V} --}} & \footnotesize{\texttt{V\textcolor{lightgray}{V} V\textcolor{lightgray}{V} V\textcolor{lightgray}{V}}} & \footnotesize{\texttt{V\textcolor{lightgray}{V} V\textcolor{lightgray}{V} --}} & \footnotesize{\texttt{V\textcolor{lightgray}{V} V\textcolor{lightgray}{V} V\textcolor{lightgray}{V}}} & \footnotesize{\texttt{V\textcolor{lightgray}{V} V\textcolor{lightgray}{V} V\textcolor{lightgray}{V}}} \\
  \footnotesize{\texttt{mathd\_algebra\_123}} & \footnotesize{\texttt{V\textcolor{lightgray}{V} \textbf{V\textcolor{lightgray}{E}} V\textcolor{lightgray}{V}}} & \footnotesize{\texttt{V\textcolor{lightgray}{V} -- V\textcolor{lightgray}{V}}} & \footnotesize{\texttt{V\textcolor{lightgray}{V} V\textcolor{lightgray}{V} V\textcolor{lightgray}{V}}} & \footnotesize{\texttt{V\textcolor{lightgray}{V} V\textcolor{lightgray}{V} V\textcolor{lightgray}{V}}} & \footnotesize{\texttt{\textbf{E\textcolor{lightgray}{V}} V\textcolor{lightgray}{V} V\textcolor{lightgray}{V}}} \\
  \footnotesize{\texttt{mathd\_algebra\_131}} & \footnotesize{\texttt{V\textcolor{lightgray}{V} V\textcolor{lightgray}{V} V\textcolor{lightgray}{V}}} & \footnotesize{\texttt{V\textcolor{lightgray}{V} V\textcolor{lightgray}{V} V\textcolor{lightgray}{V}}} & \footnotesize{\texttt{-- -- --}} & \footnotesize{\texttt{-- -- V\textcolor{lightgray}{V}}} & \footnotesize{\texttt{\textbf{E\textcolor{lightgray}{V}} -- \textbf{E\textcolor{lightgray}{V}}}} \\
  \footnotesize{\texttt{mathd\_algebra\_149}} & \footnotesize{\texttt{-- -- --}} & \footnotesize{\texttt{-- -- --}} & \footnotesize{\texttt{-- -- --}} & \footnotesize{\texttt{-- -- --}} & \footnotesize{\texttt{E\textcolor{lightgray}{E} -- --}} \\
  \bottomrule
  \end{tabular}
  }
\end{table}

\begin{table}[h]
  \centering
  \makebox[\textwidth][c]{
  \begin{tabular}{@{}lccccc@{}}
  \toprule
  & \textbf{4o mini} & \multicolumn{4}{c}{\textbf{o1 mini}} \\
  \cmidrule(l{2mm}r{2mm}){2-2}
  \cmidrule(l{2mm}){3-6}
  \textbf{Theorems} & everything & everything & formal versions & lean version & informal description \\
  \midrule
  \footnotesize{\texttt{mathd\_algebra\_153}} & \footnotesize{\texttt{-- -- --}} & \footnotesize{\texttt{-- -- --}} & \footnotesize{\texttt{-- -- --}} & \footnotesize{\texttt{-- -- --}} & \footnotesize{\texttt{-- -- --}} \\
  \footnotesize{\texttt{mathd\_algebra\_158}} & \footnotesize{\texttt{-- -- --}} & \footnotesize{\texttt{-- -- E\textcolor{lightgray}{E}}} & \footnotesize{\texttt{-- -- --}} & \footnotesize{\texttt{-- -- --}} & \footnotesize{\texttt{V\textcolor{lightgray}{V} V\textcolor{lightgray}{V} F\textcolor{lightgray}{V}}} \\
  \footnotesize{\texttt{mathd\_algebra\_188}} & \footnotesize{\texttt{-- E\textcolor{lightgray}{E} --}} & \footnotesize{\texttt{E\textcolor{lightgray}{E} -- --}} & \footnotesize{\texttt{-- -- --}} & \footnotesize{\texttt{-- -- V\textcolor{lightgray}{V}}} & \footnotesize{\texttt{V\textcolor{lightgray}{V} V\textcolor{lightgray}{V} --}} \\
  \footnotesize{\texttt{mathd\_algebra\_192}} & \footnotesize{\texttt{-- -- V\textcolor{lightgray}{V}}} & \footnotesize{\texttt{V\textcolor{lightgray}{V} V\textcolor{lightgray}{V} V\textcolor{lightgray}{V}}} & \footnotesize{\texttt{V\textcolor{lightgray}{V} -- V\textcolor{lightgray}{V}}} & \footnotesize{\texttt{V\textcolor{lightgray}{V} -- V\textcolor{lightgray}{V}}} & \footnotesize{\texttt{-- V\textcolor{lightgray}{V} --}} \\
  \footnotesize{\texttt{mathd\_algebra\_209}} & \footnotesize{\texttt{-- -- E\textcolor{lightgray}{E}}} & \footnotesize{\texttt{V\textcolor{lightgray}{V} -- \textbf{E\textcolor{lightgray}{V}}}} & \footnotesize{\texttt{-- -- --}} & \footnotesize{\texttt{-- V\textcolor{lightgray}{V} V\textcolor{lightgray}{V}}} & \footnotesize{\texttt{V\textcolor{lightgray}{V} -- V\textcolor{lightgray}{V}}} \\
  \footnotesize{\texttt{mathd\_algebra\_263}} & \footnotesize{\texttt{V\textcolor{lightgray}{V} -- --}} & \footnotesize{\texttt{V\textcolor{lightgray}{V} V\textcolor{lightgray}{V} V\textcolor{lightgray}{V}}} & \footnotesize{\texttt{V\textcolor{lightgray}{V} V\textcolor{lightgray}{V} V\textcolor{lightgray}{V}}} & \footnotesize{\texttt{V\textcolor{lightgray}{V} V\textcolor{lightgray}{V} V\textcolor{lightgray}{V}}} & \footnotesize{\texttt{-- V\textcolor{lightgray}{V} V\textcolor{lightgray}{V}}} \\
  \footnotesize{\texttt{mathd\_algebra\_282}} & \footnotesize{\texttt{-- -- --}} & \footnotesize{\texttt{-- -- --}} & \footnotesize{\texttt{-- -- --}} & \footnotesize{\texttt{-- -- --}} & \footnotesize{\texttt{-- -- --}} \\
  \footnotesize{\texttt{mathd\_algebra\_304}} & \footnotesize{\texttt{-- \textbf{V\textcolor{lightgray}{E}} V\textcolor{lightgray}{V}}} & \footnotesize{\texttt{-- -- --}} & \footnotesize{\texttt{V\textcolor{lightgray}{V} V\textcolor{lightgray}{V} V\textcolor{lightgray}{V}}} & \footnotesize{\texttt{V\textcolor{lightgray}{V} V\textcolor{lightgray}{V} V\textcolor{lightgray}{V}}} & \footnotesize{\texttt{V\textcolor{lightgray}{V} V\textcolor{lightgray}{V} --}} \\
  \footnotesize{\texttt{mathd\_algebra\_320}} & \footnotesize{\texttt{-- -- --}} & \footnotesize{\texttt{-- -- --}} & \footnotesize{\texttt{-- -- --}} & \footnotesize{\texttt{-- -- --}} & \footnotesize{\texttt{-- -- --}} \\
  \footnotesize{\texttt{mathd\_algebra\_405}} & \footnotesize{\texttt{F\textcolor{lightgray}{V} -- --}} & \footnotesize{\texttt{-- -- --}} & \footnotesize{\texttt{-- -- V\textcolor{lightgray}{V}}} & \footnotesize{\texttt{-- -- E\textcolor{lightgray}{E}}} & \footnotesize{\texttt{-- -- F\textcolor{lightgray}{V}}} \\
  \footnotesize{\texttt{mathd\_algebra\_410}} & \footnotesize{\texttt{F\textcolor{lightgray}{V} -- --}} & \footnotesize{\texttt{F\textcolor{lightgray}{V} F\textcolor{lightgray}{V} F\textcolor{lightgray}{V}}} & \footnotesize{\texttt{F\textcolor{lightgray}{V} -- F\textcolor{lightgray}{V}}} & \footnotesize{\texttt{F\textcolor{lightgray}{V} -- --}} & \footnotesize{\texttt{-- V\textcolor{lightgray}{V} --}} \\
  \footnotesize{\texttt{mathd\_algebra\_419}} & \footnotesize{\texttt{V\textcolor{lightgray}{V} V\textcolor{lightgray}{V} V\textcolor{lightgray}{V}}} & \footnotesize{\texttt{V\textcolor{lightgray}{V} V\textcolor{lightgray}{V} V\textcolor{lightgray}{V}}} & \footnotesize{\texttt{-- V\textcolor{lightgray}{V} V\textcolor{lightgray}{V}}} & \footnotesize{\texttt{V\textcolor{lightgray}{V} V\textcolor{lightgray}{V} \textbf{V\textcolor{lightgray}{E}}}} & \footnotesize{\texttt{V\textcolor{lightgray}{V} V\textcolor{lightgray}{V} V\textcolor{lightgray}{V}}} \\
  \footnotesize{\texttt{mathd\_algebra\_440}} & \footnotesize{\texttt{V\textcolor{lightgray}{V} V\textcolor{lightgray}{V} V\textcolor{lightgray}{V}}} & \footnotesize{\texttt{V\textcolor{lightgray}{V} V\textcolor{lightgray}{V} V\textcolor{lightgray}{V}}} & \footnotesize{\texttt{V\textcolor{lightgray}{V} V\textcolor{lightgray}{V} V\textcolor{lightgray}{V}}} & \footnotesize{\texttt{V\textcolor{lightgray}{V} V\textcolor{lightgray}{V} V\textcolor{lightgray}{V}}} & \footnotesize{\texttt{V\textcolor{lightgray}{V} V\textcolor{lightgray}{V} V\textcolor{lightgray}{V}}} \\
  \footnotesize{\texttt{mathd\_algebra\_455}} & \footnotesize{\texttt{-- V\textcolor{lightgray}{V} V\textcolor{lightgray}{V}}} & \footnotesize{\texttt{V\textcolor{lightgray}{V} V\textcolor{lightgray}{V} V\textcolor{lightgray}{V}}} & \footnotesize{\texttt{V\textcolor{lightgray}{V} -- --}} & \footnotesize{\texttt{V\textcolor{lightgray}{V} -- V\textcolor{lightgray}{V}}} & \footnotesize{\texttt{-- V\textcolor{lightgray}{V} V\textcolor{lightgray}{V}}} \\
  \footnotesize{\texttt{mathd\_algebra\_487}} & \footnotesize{\texttt{V\textcolor{lightgray}{V} V\textcolor{lightgray}{V} --}} & \footnotesize{\texttt{V\textcolor{lightgray}{V} V\textcolor{lightgray}{V} --}} & \footnotesize{\texttt{V\textcolor{lightgray}{V} V\textcolor{lightgray}{V} V\textcolor{lightgray}{V}}} & \footnotesize{\texttt{V\textcolor{lightgray}{V} V\textcolor{lightgray}{V} V\textcolor{lightgray}{V}}} & \footnotesize{\texttt{F\textcolor{lightgray}{V} \textbf{E\textcolor{lightgray}{V}} F\textcolor{lightgray}{V}}} \\
  \footnotesize{\texttt{mathd\_algebra\_493}} & \footnotesize{\texttt{V\textcolor{lightgray}{V} V\textcolor{lightgray}{V} V\textcolor{lightgray}{V}}} & \footnotesize{\texttt{-- V\textcolor{lightgray}{V} V\textcolor{lightgray}{V}}} & \footnotesize{\texttt{-- V\textcolor{lightgray}{V} V\textcolor{lightgray}{V}}} & \footnotesize{\texttt{V\textcolor{lightgray}{V} -- --}} & \footnotesize{\texttt{V\textcolor{lightgray}{V} V\textcolor{lightgray}{V} V\textcolor{lightgray}{V}}} \\
  \footnotesize{\texttt{mathd\_algebra\_509}} & \footnotesize{\texttt{-- V\textcolor{lightgray}{V} V\textcolor{lightgray}{V}}} & \footnotesize{\texttt{V\textcolor{lightgray}{V} -- V\textcolor{lightgray}{V}}} & \footnotesize{\texttt{V\textcolor{lightgray}{V} V\textcolor{lightgray}{V} V\textcolor{lightgray}{V}}} & \footnotesize{\texttt{-- V\textcolor{lightgray}{V} V\textcolor{lightgray}{V}}} & \footnotesize{\texttt{-- V\textcolor{lightgray}{V} V\textcolor{lightgray}{V}}} \\
  \footnotesize{\texttt{mathd\_algebra\_513}} & \footnotesize{\texttt{V\textcolor{lightgray}{V} -- --}} & \footnotesize{\texttt{-- V\textcolor{lightgray}{V} V\textcolor{lightgray}{V}}} & \footnotesize{\texttt{V\textcolor{lightgray}{V} V\textcolor{lightgray}{V} V\textcolor{lightgray}{V}}} & \footnotesize{\texttt{V\textcolor{lightgray}{V} V\textcolor{lightgray}{V} V\textcolor{lightgray}{V}}} & \footnotesize{\texttt{V\textcolor{lightgray}{V} V\textcolor{lightgray}{V} V\textcolor{lightgray}{V}}} \\
  \footnotesize{\texttt{mathd\_numbertheory\_22}} & \footnotesize{\texttt{E\textcolor{lightgray}{E} -- --}} & \footnotesize{\texttt{F\textcolor{lightgray}{V} F\textcolor{lightgray}{V} F\textcolor{lightgray}{V}}} & \footnotesize{\texttt{F\textcolor{lightgray}{V} F\textcolor{lightgray}{V} F\textcolor{lightgray}{V}}} & \footnotesize{\texttt{-- -- --}} & \footnotesize{\texttt{-- E\textcolor{lightgray}{E} --}} \\
  \footnotesize{\texttt{mathd\_numbertheory\_24}} & \footnotesize{\texttt{-- -- --}} & \footnotesize{\texttt{V\textcolor{lightgray}{V} V\textcolor{lightgray}{V} V\textcolor{lightgray}{V}}} & \footnotesize{\texttt{-- -- --}} & \footnotesize{\texttt{V\textcolor{lightgray}{V} -- --}} & \footnotesize{\texttt{-- V\textcolor{lightgray}{V} --}} \\
  \footnotesize{\texttt{mathd\_numbertheory\_42}} & \footnotesize{\texttt{V\textcolor{lightgray}{V} -- V\textcolor{lightgray}{V}}} & \footnotesize{\texttt{-- V\textcolor{lightgray}{V} V\textcolor{lightgray}{V}}} & \footnotesize{\texttt{-- -- --}} & \footnotesize{\texttt{-- -- --}} & \footnotesize{\texttt{-- E\textcolor{lightgray}{E} --}} \\
  \footnotesize{\texttt{mathd\_numbertheory\_45}} & \footnotesize{\texttt{V\textcolor{lightgray}{V} V\textcolor{lightgray}{V} --}} & \footnotesize{\texttt{-- -- V\textcolor{lightgray}{V}}} & \footnotesize{\texttt{-- V\textcolor{lightgray}{V} V\textcolor{lightgray}{V}}} & \footnotesize{\texttt{-- -- V\textcolor{lightgray}{V}}} & \footnotesize{\texttt{-- V\textcolor{lightgray}{V} V\textcolor{lightgray}{V}}} \\
  \footnotesize{\texttt{mathd\_numbertheory\_64}} & \footnotesize{\texttt{-- -- --}} & \footnotesize{\texttt{-- V\textcolor{lightgray}{V} V\textcolor{lightgray}{V}}} & \footnotesize{\texttt{V\textcolor{lightgray}{V} -- --}} & \footnotesize{\texttt{-- -- --}} & \footnotesize{\texttt{-- V\textcolor{lightgray}{V} V\textcolor{lightgray}{V}}} \\
  \footnotesize{\texttt{mathd\_numbertheory\_127}} & \footnotesize{\texttt{-- F\textcolor{lightgray}{V} V\textcolor{lightgray}{V}}} & \footnotesize{\texttt{V\textcolor{lightgray}{V} -- F\textcolor{lightgray}{V}}} & \footnotesize{\texttt{V\textcolor{lightgray}{V} V\textcolor{lightgray}{V} --}} & \footnotesize{\texttt{-- F\textcolor{lightgray}{V} --}} & \footnotesize{\texttt{F\textcolor{lightgray}{V} F\textcolor{lightgray}{V} F\textcolor{lightgray}{V}}} \\
  \footnotesize{\texttt{mathd\_numbertheory\_149}} & \footnotesize{\texttt{-- -- --}} & \footnotesize{\texttt{-- -- --}} & \footnotesize{\texttt{-- -- --}} & \footnotesize{\texttt{-- -- --}} & \footnotesize{\texttt{-- -- --}} \\
  \footnotesize{\texttt{mathd\_numbertheory\_150}} & \footnotesize{\texttt{-- -- --}} & \footnotesize{\texttt{-- -- --}} & \footnotesize{\texttt{-- -- --}} & \footnotesize{\texttt{-- -- --}} & \footnotesize{\texttt{-- -- --}} \\
  \footnotesize{\texttt{mathd\_numbertheory\_175}} & \footnotesize{\texttt{V\textcolor{lightgray}{V} V\textcolor{lightgray}{V} V\textcolor{lightgray}{V}}} & \footnotesize{\texttt{-- V\textcolor{lightgray}{V} --}} & \footnotesize{\texttt{-- V\textcolor{lightgray}{V} V\textcolor{lightgray}{V}}} & \footnotesize{\texttt{-- V\textcolor{lightgray}{V} V\textcolor{lightgray}{V}}} & \footnotesize{\texttt{V\textcolor{lightgray}{V} V\textcolor{lightgray}{V} V\textcolor{lightgray}{V}}} \\
  \footnotesize{\texttt{mathd\_numbertheory\_185}} & \footnotesize{\texttt{-- V\textcolor{lightgray}{V} V\textcolor{lightgray}{V}}} & \footnotesize{\texttt{V\textcolor{lightgray}{V} V\textcolor{lightgray}{V} --}} & \footnotesize{\texttt{V\textcolor{lightgray}{V} V\textcolor{lightgray}{V} V\textcolor{lightgray}{V}}} & \footnotesize{\texttt{V\textcolor{lightgray}{V} -- --}} & \footnotesize{\texttt{V\textcolor{lightgray}{V} V\textcolor{lightgray}{V} --}} \\
  \footnotesize{\texttt{mathd\_numbertheory\_188}} & \footnotesize{\texttt{V\textcolor{lightgray}{V} V\textcolor{lightgray}{V} --}} & \footnotesize{\texttt{V\textcolor{lightgray}{V} -- --}} & \footnotesize{\texttt{V\textcolor{lightgray}{V} V\textcolor{lightgray}{V} --}} & \footnotesize{\texttt{-- -- --}} & \footnotesize{\texttt{-- V\textcolor{lightgray}{V} --}} \\
  \footnotesize{\texttt{mathd\_numbertheory\_207}} & \footnotesize{\texttt{V\textcolor{lightgray}{V} V\textcolor{lightgray}{V} V\textcolor{lightgray}{V}}} & \footnotesize{\texttt{-- -- V\textcolor{lightgray}{V}}} & \footnotesize{\texttt{V\textcolor{lightgray}{V} V\textcolor{lightgray}{V} --}} & \footnotesize{\texttt{V\textcolor{lightgray}{V} -- V\textcolor{lightgray}{V}}} & \footnotesize{\texttt{\textbf{V\textcolor{lightgray}{E}} V\textcolor{lightgray}{V} V\textcolor{lightgray}{V}}} \\
  \footnotesize{\texttt{mathd\_numbertheory\_212}} & \footnotesize{\texttt{V\textcolor{lightgray}{V} V\textcolor{lightgray}{V} V\textcolor{lightgray}{V}}} & \footnotesize{\texttt{-- V\textcolor{lightgray}{V} V\textcolor{lightgray}{V}}} & \footnotesize{\texttt{-- V\textcolor{lightgray}{V} V\textcolor{lightgray}{V}}} & \footnotesize{\texttt{V\textcolor{lightgray}{V} V\textcolor{lightgray}{V} V\textcolor{lightgray}{V}}} & \footnotesize{\texttt{V\textcolor{lightgray}{V} V\textcolor{lightgray}{V} V\textcolor{lightgray}{V}}} \\
  \footnotesize{\texttt{mathd\_numbertheory\_221}} & \footnotesize{\texttt{-- -- --}} & \footnotesize{\texttt{-- -- --}} & \footnotesize{\texttt{-- -- --}} & \footnotesize{\texttt{-- -- --}} & \footnotesize{\texttt{-- -- --}} \\
  \footnotesize{\texttt{mathd\_numbertheory\_234}} & \footnotesize{\texttt{V\textcolor{lightgray}{V} -- --}} & \footnotesize{\texttt{V\textcolor{lightgray}{V} V\textcolor{lightgray}{V} --}} & \footnotesize{\texttt{-- -- V\textcolor{lightgray}{V}}} & \footnotesize{\texttt{V\textcolor{lightgray}{V} -- V\textcolor{lightgray}{V}}} & \footnotesize{\texttt{V\textcolor{lightgray}{V} V\textcolor{lightgray}{V} --}} \\
  \footnotesize{\texttt{mathd\_numbertheory\_252}} & \footnotesize{\texttt{-- -- --}} & \footnotesize{\texttt{-- V\textcolor{lightgray}{V} V\textcolor{lightgray}{V}}} & \footnotesize{\texttt{V\textcolor{lightgray}{V} V\textcolor{lightgray}{V} --}} & \footnotesize{\texttt{-- -- --}} & \footnotesize{\texttt{V\textcolor{lightgray}{V} V\textcolor{lightgray}{V} --}} \\
  \footnotesize{\texttt{mathd\_numbertheory\_293}} & \footnotesize{\texttt{-- -- --}} & \footnotesize{\texttt{-- -- --}} & \footnotesize{\texttt{-- -- --}} & \footnotesize{\texttt{-- -- --}} & \footnotesize{\texttt{\textbf{E\textcolor{lightgray}{V}} V\textcolor{lightgray}{V} --}} \\
  \footnotesize{\texttt{mathd\_numbertheory\_296}} & \footnotesize{\texttt{E\textcolor{lightgray}{E} F\textcolor{lightgray}{V} --}} & \footnotesize{\texttt{E\textcolor{lightgray}{E} V\textcolor{lightgray}{V} F\textcolor{lightgray}{V}}} & \footnotesize{\texttt{-- V\textcolor{lightgray}{V} --}} & \footnotesize{\texttt{-- -- --}} & \footnotesize{\texttt{V\textcolor{lightgray}{V} V\textcolor{lightgray}{V} --}} \\
  \footnotesize{\texttt{mathd\_numbertheory\_299}} & \footnotesize{\texttt{V\textcolor{lightgray}{V} V\textcolor{lightgray}{V} V\textcolor{lightgray}{V}}} & \footnotesize{\texttt{-- -- V\textcolor{lightgray}{V}}} & \footnotesize{\texttt{-- V\textcolor{lightgray}{V} V\textcolor{lightgray}{V}}} & \footnotesize{\texttt{V\textcolor{lightgray}{V} V\textcolor{lightgray}{V} V\textcolor{lightgray}{V}}} & \footnotesize{\texttt{V\textcolor{lightgray}{V} V\textcolor{lightgray}{V} V\textcolor{lightgray}{V}}} \\
  \footnotesize{\texttt{mathd\_numbertheory\_321}} & \footnotesize{\texttt{\textbf{V\textcolor{lightgray}{E}} \textbf{V\textcolor{lightgray}{E}} \textbf{V\textcolor{lightgray}{E}}}} & \footnotesize{\texttt{V\textcolor{lightgray}{V} -- V\textcolor{lightgray}{V}}} & \footnotesize{\texttt{-- V\textcolor{lightgray}{V} V\textcolor{lightgray}{V}}} & \footnotesize{\texttt{-- \textbf{E\textcolor{lightgray}{V}} --}} & \footnotesize{\texttt{F\textcolor{lightgray}{V} -- --}} \\
  \footnotesize{\texttt{mathd\_numbertheory\_328}} & \footnotesize{\texttt{V\textcolor{lightgray}{V} V\textcolor{lightgray}{V} V\textcolor{lightgray}{V}}} & \footnotesize{\texttt{-- -- V\textcolor{lightgray}{V}}} & \footnotesize{\texttt{-- V\textcolor{lightgray}{V} V\textcolor{lightgray}{V}}} & \footnotesize{\texttt{V\textcolor{lightgray}{V} V\textcolor{lightgray}{V} --}} & \footnotesize{\texttt{V\textcolor{lightgray}{V} -- V\textcolor{lightgray}{V}}} \\
  \footnotesize{\texttt{mathd\_numbertheory\_342}} & \footnotesize{\texttt{V\textcolor{lightgray}{V} -- V\textcolor{lightgray}{V}}} & \footnotesize{\texttt{-- -- --}} & \footnotesize{\texttt{V\textcolor{lightgray}{V} V\textcolor{lightgray}{V} --}} & \footnotesize{\texttt{-- V\textcolor{lightgray}{V} V\textcolor{lightgray}{V}}} & \footnotesize{\texttt{-- V\textcolor{lightgray}{V} --}} \\
  \footnotesize{\texttt{mathd\_numbertheory\_543}} & \footnotesize{\texttt{-- -- --}} & \footnotesize{\texttt{-- -- --}} & \footnotesize{\texttt{-- -- --}} & \footnotesize{\texttt{-- -- --}} & \footnotesize{\texttt{-- -- --}} \\
  \footnotesize{\texttt{mathd\_numbertheory\_552}} & \footnotesize{\texttt{-- -- --}} & \footnotesize{\texttt{-- -- --}} & \footnotesize{\texttt{-- -- --}} & \footnotesize{\texttt{-- -- --}} & \footnotesize{\texttt{\textbf{E\textcolor{lightgray}{V}} E\textcolor{lightgray}{E} --}} \\
  \footnotesize{\texttt{mathd\_numbertheory\_629}} & \footnotesize{\texttt{-- -- --}} & \footnotesize{\texttt{-- -- --}} & \footnotesize{\texttt{V\textcolor{lightgray}{V} V\textcolor{lightgray}{V} --}} & \footnotesize{\texttt{-- -- --}} & \footnotesize{\texttt{-- -- \textbf{F\textcolor{lightgray}{E}}}} \\
  \footnotesize{\texttt{mathd\_numbertheory\_640}} & \footnotesize{\texttt{V\textcolor{lightgray}{V} V\textcolor{lightgray}{V} V\textcolor{lightgray}{V}}} & \footnotesize{\texttt{V\textcolor{lightgray}{V} V\textcolor{lightgray}{V} --}} & \footnotesize{\texttt{-- V\textcolor{lightgray}{V} --}} & \footnotesize{\texttt{V\textcolor{lightgray}{V} V\textcolor{lightgray}{V} --}} & \footnotesize{\texttt{V\textcolor{lightgray}{V} V\textcolor{lightgray}{V} V\textcolor{lightgray}{V}}} \\
  \footnotesize{\texttt{mathd\_numbertheory\_765}} & \footnotesize{\texttt{F\textcolor{lightgray}{V} F\textcolor{lightgray}{V} F\textcolor{lightgray}{V}}} & \footnotesize{\texttt{-- V\textcolor{lightgray}{V} \textbf{V\textcolor{lightgray}{E}}}} & \footnotesize{\texttt{-- -- --}} & \footnotesize{\texttt{-- -- F\textcolor{lightgray}{V}}} & \footnotesize{\texttt{F\textcolor{lightgray}{V} V\textcolor{lightgray}{V} --}} \\
  \footnotesize{\texttt{numbertheory\_2pownm...}} & \footnotesize{\texttt{-- -- --}} & \footnotesize{\texttt{-- -- --}} & \footnotesize{\texttt{-- -- --}} & \footnotesize{\texttt{-- -- --}} & \footnotesize{\texttt{-- -- --}} \\
  \footnotesize{\texttt{numbertheory\_exk2po...}} & \footnotesize{\texttt{V\textcolor{lightgray}{V} -- --}} & \footnotesize{\texttt{V\textcolor{lightgray}{V} -- --}} & \footnotesize{\texttt{-- V\textcolor{lightgray}{V} V\textcolor{lightgray}{V}}} & \footnotesize{\texttt{V\textcolor{lightgray}{V} -- V\textcolor{lightgray}{V}}} & \footnotesize{\texttt{-- -- --}} \\
  \footnotesize{\texttt{numbertheory\_fxeq4p...}} & \footnotesize{\texttt{-- -- --}} & \footnotesize{\texttt{-- -- --}} & \footnotesize{\texttt{-- -- --}} & \footnotesize{\texttt{-- -- --}} & \footnotesize{\texttt{-- V\textcolor{lightgray}{V} V\textcolor{lightgray}{V}}} \\
  \footnotesize{\texttt{numbertheory\_notequ...}} & \footnotesize{\texttt{-- -- --}} & \footnotesize{\texttt{-- -- --}} & \footnotesize{\texttt{V\textcolor{lightgray}{V} -- V\textcolor{lightgray}{V}}} & \footnotesize{\texttt{-- -- --}} & \footnotesize{\texttt{-- -- \textbf{E\textcolor{lightgray}{V}}}} \\
  \footnotesize{\texttt{numbertheory\_sumkmu...}} & \footnotesize{\texttt{-- -- --}} & \footnotesize{\texttt{-- -- --}} & \footnotesize{\texttt{-- -- --}} & \footnotesize{\texttt{-- -- V\textcolor{lightgray}{V}}} & \footnotesize{\texttt{-- -- --}} \\
  \bottomrule
  \end{tabular}
  }
\end{table}

\end{document}